\documentclass{llncs}

\usepackage{amsmath}
\usepackage{amsfonts}
\usepackage{amssymb}

\usepackage{graphicx}
\usepackage{epic}
\usepackage{eepic}
\usepackage{verbatim}
\renewcommand{\epsilon}{\varepsilon}

\newcommand{\ol}{\overline}

\newcommand{\emp}{\emptyset}
\newcommand{\Sig}{\Sigma}

\newcommand{\bi}{\begin{itemize}}
\newcommand{\ei}{\end{itemize}}
\newcommand{\be}{\begin{enumerate}}
\newcommand{\ee}{\end{enumerate}}
\newcommand{\bd}{\begin{description}}
\newcommand{\ed}{\end{description}}
\newcommand{\bq}{\begin{quote}}
\newcommand{\eq}{\end{quote}}
\newcommand{\txt}[1]{\mbox{ #1 }}

\newcommand{\etc}{\mbox{\it etc.}}
\newcommand{\ie}{\mbox{\it i.e.}}

\def\shu{\mathbin{\mathchoice
{\rule{.3pt}{1ex}\rule{.3em}{.3pt}\rule{.3pt}{1ex}
\rule{.3em}{.3pt}\rule{.3pt}{1ex}}
{\rule{.3pt}{1ex}\rule{.3em}{.3pt}\rule{.3pt}{1ex}
\rule{.3em}{.3pt}\rule{.3pt}{1ex}}
{\rule{.2pt}{.7ex}\rule{.2em}{.2pt}\rule{.2pt}{.7ex}
\rule{.2em}{.2pt}\rule{.2pt}{.7ex}}
{\rule{.3pt}{1ex}\rule{.3em}{.3pt}\rule{.3pt}{1ex}
\rule{.3em}{.3pt}\rule{.3pt}{1ex}}\mkern2mu}}

\newcommand{\cC}{{\mathcal C}}

\newcommand{\cL}{{\mathcal L}}

\newcommand{\cN}{{\mathcal N}}

\newcommand{\rhoR}{R}

\title{Quotient  Complexity of Closed Languages
\thanks{This work was supported by the
Natural Sciences and Engineering Research Council of Canada grant OGP0000871
and by VEGA grant 2/0111/09.
}}

\author{Janusz~Brzozowski\inst{1} \and Galina Jir\'askov\'a\inst{2} \and Chenglong Zou \inst{1}}

\authorrunning{Brzozowski, Jir\'askov\'a, Zou}

\institute{David R. Cheriton School of Computer Science, University of Waterloo, \\
Waterloo, ON, Canada N2L 3G1\\
\{{\tt brzozo@,c2zou@student.math.\}uwaterloo.ca} \quad
\and
Mathematical Institute,
Slovak Academy of Science,\\
Gre\v{s}\'akova 6, 040 01 Ko\v{s}ice, Slovakia\\
\{{\tt jiraskov@saske.sk}\}
}

\begin{document}

\maketitle

\begin{abstract}
\label{***abstr}

A language $L$ is prefix-closed if, whenever a word $w$ is in~$L$, 
then every prefix of $w$ is also in $L$. 
We define suffix-, factor-, and subword-closed  languages in the same way, 
where by subword we mean subsequence.
We study the quotient complexity (usually called state complexity) 
of operations on prefix-, suffix-, factor-, and subword-closed languages.
We find tight upper bounds on the complexity 
of the prefix-, suffix-, factor-, and subword-closure of arbitrary languages, 
and on the complexity of  boolean operations, concatenation, star and reversal 
in each of the four classes of closed languages. 
We show that repeated application of
positive closure and complement to a closed language
results in at most four distinct languages, 
while Kleene closure and complement gives at most eight languages.
\medskip

\end{abstract}

\section{Introduction}
\label{***intro}

The \emph{state complexity of a regular language} $L$ 
is the number of states in the minimal deterministic finite automaton (dfa) recognizing $L$. 
The \emph{state complexity of an operation}
$f(K,L)$ (or $g(L)$) in a subclass $\cC$ of regular languages 
is the maximal state complexity of the language $f(K,L)$ (or $g(L)$),
when $K$ and $L$ range over all languages in $\cC$.
For a detailed discussion of general issues of state complexity 
see~\cite{Brz09,Yu01} and the reference lists in those papers.
In 1994  the complexity of concatenation, star, left and right quotients, 
reversal, intersection and union in  regular languages 
were examined in detail in~\cite{YZS94}.
The complexity of operations was also considered 
in several subclasses of regular languages:
finite~\cite{Yu01}, unary~\cite{PiSh02,YZS94},   prefix-free~\cite{HaSa09} and suffix-free~\cite{HaSa09a}, 
and ideal languages~\cite{BJL09}. 
These studies  show that the complexity can be significantly lower 
in a subclass than in the general case.
Here we examine state complexity in the classes of
prefix-, suffix-, factor-, and subword-closed regular languages.

There are several reasons for considering closed languages. 
They appear often in theoretical computer science.
Subword-closed languages were studied in 1969~\cite{Hai69}, 
and also in 1973~\cite{Thi73}.
Suffix-closed languages were considered in 1974~\cite{GiKo74}, 
and later in~\cite{GaSi76,HSY01,VeGi79}.
Factor-closed languages, also called \emph{factorial},  
have received some attention, for example, in~\cite{AvFr05,LuVa90}.
Subword-closed languages were studied in~\cite{Okh07}.
Prefix-closed languages play a role in predictable semiautomata~\cite{BrSa09}.
All four classes of closed languages  were examined in~\cite{AnBr09}, 
and decision problems for closed languages were studied in~\cite{BSX09}.
A language is a \emph{left ideal} 
(respectively, \emph{right, two-sided}, \emph{all-sided ideal}) 
if $L=\Sig^*L$, (respectively, $L=L\Sig^*$, $L=\Sig^*L\Sig^*$ and $L=\Sig^*\shu L$),  where $\Sig^*\shu L$ is the shuffle of $\Sig^*$ with $L$).
Closed languages are related to ideal languages   as follows~\cite{AnBr09}:  
For every non-empty $L$,
$L$ is a right (left, two-sided, all-sided) ideal, 
if and only if $\overline{L}$ is a prefix(suffix, factor, subword)-closed language.
Closed languages are defined by binary relations  
``is a prefix of'' (respectively,
``is a suffix of", ``is a factor of'', ``is a subword of'')~\cite{AnBr09}, 
and are special cases of convex languages~\cite{AnBr09,Thi73}.
The fact that the four classes of closed languages 
are related to each other permits us 
to obtain many complexity results using similar methods.

\section{Quotient Complexity}
\label{***prelim}

If $\Sigma$ is a non-empty finite \emph{alphabet}, 
then $\Sigma^*$ is the free monoid generated by~$\Sigma$. 
A \emph{word} is any element of $\Sigma^*$, 
and $\epsilon$ is the \emph{empty word}. 
The length of a word $w\in \Sigma^*$ is $|w|$.
A \emph{language} over $\Sigma$ is any subset of $\Sigma^*$.
The cardinality of a set is denoted by $|S|$.

If $w=uxv$ for some $u,v,x\in\Sigma^*$, 
then $u$ is a \emph{prefix} of $w$, $v$ is a \emph{suffix} of $w$, 
and $x$ is a \emph{factor} of $w$.
If $w=w_0a_1w_1\cdots a_nw_n$, 
where $a_1,\ldots,a_n\in \Sigma$, and $w_0,\ldots,w_n\in\Sigma^*$, 
then $v=a_1\cdots a_n$ is a \emph{subword}  of $w$.

A~language $L$ is \emph{prefix-closed} 
if  $w \in L$ implies that every prefix of $w$ is also in~$L$.
In the same way, we define  
\emph{suffix-}, \emph{factor-}, and \emph{subword-closed} languages. 
A language is \emph{closed} if it is prefix-, suffix-, factor-, or subword-closed.

The following set operations are defined on languages: 
\emph{complement} ($\ol{L}=\Sigma^*\setminus L$),  
\emph{union}  ($K\cup L$),  
\emph{intersection} ($K\cap L$),  
\emph{difference} ($K\setminus L$), and 
\emph{symmetric difference} ($K\oplus L$).
A general \emph{boolean operation} with two arguments is denoted by $K\circ L$.
We also  define the \emph{product}, 
usually called  \emph{concatenation} or \emph{catenation},   
($KL=\{w\in \Sigma^*\mid w=uv, u\in K, v\in L\}$),   
(Kleene) \emph{star} ($K^*=\bigcup_{i\ge 0}K^i$), and 
\emph{positive closure} ($K^+=\bigcup_{i\ge 1}K^i$).
The \emph{reverse} $w^R$ of a word $w\in\Sigma^*$ is defined as follows: $\epsilon^R=\epsilon$, and $(wa)^R=aw^R$.
The \emph{reverse} of a language $L$ is denoted by $L^\rhoR$ and is defined as
$L^R=\{w^R\mid w\in L\}$.

\emph{Regular languages} over $\Sigma$ 
are languages that can be obtained from the \emph{set of basic languages} $\{\emp,\{\epsilon\}\} \cup \{\{a\}\mid  a\in \Sigma\}$,  
using a finite number of operations of union, product  and  star. 
Such languages are usually denoted by regular expressions. 
If $E$ is a regular expression, 
then $\cL(E)$ is the language denoted by that expression. 
For example, $E=(\epsilon\cup a)^* b$ denotes $L=\cL(E)=(\{\epsilon\}\cup\{a\})^*\{b\}$. 
We usually do not distinguish notationally 
between regular languages and regular expressions; 
the meaning is clear from the context. 

A \emph{deterministic finite automaton} (dfa) is a tuple
 $\mathcal{D}=(Q,\Sigma,\delta,q_0,F)$, where 
 $Q$ is a set of \emph{states}, 
 $\Sigma$ is the \emph{alphabet}, 
 $\delta:Q\times\Sigma\rightarrow Q$ is the \emph{transition function},
 $q_0$ is the \emph{initial state}, and 
 $F$ is the set of \emph{final} or \emph{accepting states}.
A \emph{nondeterministic finite automaton}  (nfa) is a tuple
 $\mathcal{N}=(Q,\Sigma,\eta,Q_0,F)$, where 
 $Q$, $\Sigma$ and $F$ are as in a dfa, 
 $\eta:Q\times\Sigma\rightarrow 2^{Q}$ is the \emph{transition function} and
 $Q_0\subseteq Q$ is the \emph{set of initial states}. 
If $\eta$ also allows $\epsilon$, 
\ie, $\eta:Q\times(\Sigma\cup\{\epsilon\})\rightarrow 2^{Q}$, 
we call $\cN$ an \emph{$\epsilon$-nfa}.

Our approach to quotient complexity follows closely that of~\cite{Brz09}.
Since state complexity  is a property of a language, 
it is more appropriately defined in language-theoretic terms.
The \emph{left quotient}, or simply \emph{quotient,} of a language $L$ by a word $w$ is  the language $L_w=\{x\in \Sigma^*\mid wx\in L \}$.
The \emph{quotient complexity} of $L$ 
is the number of distinct quotients of $L$, and is denoted by $\kappa(L)$.

Quotients of regular languages~\cite{Brz64,Brz09} 
can be computed as follows:
First, the \emph{$\epsilon$-function}  $L^\epsilon$ of a regular language $L$ is
 $L^\epsilon=\emptyset$ if $\epsilon\not\in L$ and
 $L^\epsilon=\epsilon$ if $\epsilon\in L$.
The {quotient by a letter} $a\in\Sigma$ 
is computed by structural induction:
 $b_a=\emptyset$ if $b\in \{\emptyset,\epsilon\}$ or $b\in\Sigma$ and $b\not= a$, and
 $b_a=\epsilon$ if $b=a$;
 $(\overline{L})_a =\overline{L_a};\,
 (K\cup L)_a = K_a\cup L_a; \,
 (KL)_a = K_aL \cup K^\epsilon L_a;\,
 (K^*)_a = K_aK^*$.
The quotient by a word $w\in\Sigma^*$  is computed by
induction on the length of $w$:
 $L_\epsilon =  L;\,
 L_w = L_a $  if $w=a\in \Sigma$;\,
 $L_{wa} = (L_w)_a$.
A quotient $L_w$ is \emph{accepting} if $\epsilon\in L_w$; 
otherwise it is \emph{rejecting}.

The \emph{quotient automaton} of a regular language $L$ is
 $\mathcal{D}=(Q, \Sigma, \delta, q_0,F)$, where 
 $Q=\{L_w\mid w\in\Sigma^*\}$, 
 $\delta(L_w,a)=L_{wa}$,
 $q_0=L_\epsilon=L$, and 
 $F=\{L_w\mid (L_w)^\epsilon=\epsilon\}$.
This is the minimal dfa accepting $L$; 
hence quotient complexity of $L$ is equal to the state complexity of $L$. 
However, there are some advantages to using quotients~\cite{Brz09}. 
If  a language $L$ has the empty quotient, we say that $L$ \emph{has}~$\emptyset$.

To simplify the notation, 
we write $(L_w)^\epsilon$ as $L_w^\epsilon$.
Whenever convenient, the following formulas are used 
to establish upper bounds on quotient complexity:

\begin{proposition}[\cite{Brz64,Brz09}]\label{prop}       
 If $K$ and $L$ are regular languages, then
 \begin{equation}\label{eq:bool}
  (\overline{L})_w = \overline{L_w}; \quad (K\circ L)_w=K_w\circ L_w.
 \end{equation}
 \begin{equation}\label{eq:prod}
  (KL)_w =  K_wL \cup  K^\epsilon L_w\cup
          \left(\bigcup_{{w=uv}\atop {\;\; u,v\in\Sigma^+}} K_u^\epsilon L_v\right).
 \end{equation}
 \begin{equation}\label{eq:star}
  (L^*)_\epsilon=\epsilon\cup LL^*, \quad 
  (L^*)_w=  \left( L_w \cup \bigcup_{{w=uv}\atop {\;\; u,v\in\Sigma^+}}
           (L^*)_u^\epsilon L_v \right) L^* \;\mbox{  for } w\in\Sigma^+.
 \end{equation}
\end{proposition}

\section{Closure Operations}
\label{***closure}

We now turn to the closure of languages under binary relations.
All the relations that we study in this paper are partial orders.
Let $\unlhd$ be a partial order on $\Sigma^*$;   
the \emph{$\unlhd$-closure} of a language $L$ is the language 
$_{\unlhd}L=\{x\in\Sigma^*\mid x \, \unlhd\, w\textrm{ for some }w\in L\}$. 
We use $\leq,\;\preceq,\;\sqsubseteq,\;\Subset$ for the relations 
``is a prefix of'', ``is a suffix of'', ``is a factor of'', ``is a subword of'', respectively.

Suppose $L$ is an arbitrary regular language of complexity $n$. 
If $n=1$ then $L=\emptyset$ or $L=\Sigma^*$,
and each closure is $L$.
We show that the worst-case complexity 
for prefix-closure is $n$,  
for suffix-closure it is $2^n-1$, and 
for factor-closure it is $2^{n-1}$. 
These bounds are tight for binary languages.
Subword-closure of languages was previously studied by Okhotin~\cite{Okh07} 
under the name ``scattered subwords'', 
but tight upper bounds were not established. 
Our next theorem solves this problem.

\begin{theorem}[Closure Operations]\label{thm:CLOSURE}
 Let $L$ be a regular language with $\kappa(L)=n\geq 2$.
 Let $_{\leq}L,\;_{\preceq}L,\;_{\sqsubseteq}L,\;_{\Subset}L$
 be the prefix-closure, suffix-closure, factor-closure,
 and subword-closure of $L$, respectively. Then\\
 1. $\kappa(_{\leq}L)\leq n$.\\
 2. $\kappa(_{\preceq}L)\leq 2^n-1$  if $L$ does not have $\emptyset$,
    and  $\kappa(_{\preceq}L)\leq 2^{n-1}$ otherwise.\\
 3. $\kappa({}_{\sqsubseteq}L)\leq 2^{n-1}$.\\
 4. $\kappa(_{\Subset}L)\leq 2^{n-2}+1$.\\
 The last bound is tight if $|\Sigma|\ge n-2$;
 the other bounds are tight if $|\Sigma|\geq 2$.
\end{theorem}

\begin{proof}
 1. Given a language $L$ recognized by dfa $\mathcal{D}$, 
 to get the dfa for its prefix-closure $_\le L$, 
 we need only  make each non-empty state accepting.
 Hence $\kappa({_\le}L)\le n$.
 For tightness, consider the language
 $L=\{a^i\mid i\leq n-2\}$.
 We have ${_\leq}L=L$ and  $\kappa({_\leq}L)=n$.

 2. Having a quotient automaton of a language $L$,
 we can construct an nfa for its suffix-closure 
 by making each non-empty state  initial.
 The equivalent dfa has at most $2^n-1$ states
 if $L$ does not have the empty quotient (the empty set of states cannot be reached),
 and at most $2^{n-1}$ states otherwise.
 To prove tightness, consider the language $L$
 defined by the quotient automaton shown in Fig.~\ref{fig:suffix1}.
 Construct an nfa for the suffix-closure of $L$,
 by making all states initial.
 Let us show that the corresponding subset automaton
 has $2^n-1$ reachable and pairwise inequivalent states.
 \begin{figure}[hbt]
    \centerline{\includegraphics[scale=0.40]{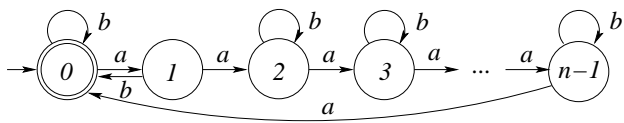}}
    \caption{Quotient automaton of a language $L$ which does not have $\emptyset$.}
    \label{fig:suffix1}
 \end{figure}

 We prove reachability by induction on the size of subsets.
 The basis, $|S|=n$, holds true since $\{0,1,\ldots,n-1\}$
 is the initial state.
 Assume that each set of size $k$ is reachable,
 and let $S$ be a set of size $k-1$.
 If $S$ contains state $0$ but does not contain state $1$,
 then it can be reached from the set $S\cup \{1\}$ of size $k$ by $b$.
 If $S$ contains both $0$ and $1$, then there is a state $i$
 such that $i\in S$ and $i+1\notin S$.
 Then $S$ can be reached from  $\{s-i \bmod n \mid s\in S\}$  by $a^i$. 
 The latter set contains $0$ and does not contain $1$,
 and so is reachable.
 If a non-empty  $S$ does not contain  $0$,
 then it can be reached from  $\{s-\min S\mid s\in S\}$,
 which contains  $0$, by $a^{\min S}$.

 To prove inequivalence notice that
 the word $a^{n-i}$
 is accepted by the nfa only from state $i$
 for all $i=0,1,\ldots,n-1$.
 It turns out that all the states in the subset automaton are pairwise inequivalent.

 Now consider  the case where a language has $\emptyset$.
 Let $L$ be the language defined
 by the quotient automaton shown in Fig.~\ref{fig:suffix2}.
 We first remove state $n-1$ and all transitions going to this state,
 and then construct an nfa  as above.
 The proof of reachability of 
 all non-empty subsets of $\{0,1,\ldots,n-2\}$ is the same as in the previous case.
 The empty set can be reached from $\{0\}$ by $b$.
 For inequivalence, 
 $(ab)^n$ is accepted 
 only from  $0$,  and 
 $a^{n-1-i}(ab)^n$ only from  $i$  for $i=1,2,\ldots,n-2$.

\begin{figure}[hbt]
    \centerline{\includegraphics[scale=0.40]{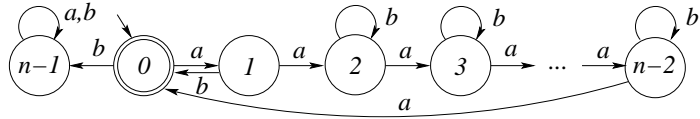}}
    \caption{Quotient automaton of a language $L$ which has $\emptyset$.}
    \label{fig:suffix2}
 \end{figure}

 3. Suppose we have the quotient automaton of a language $L$.
 To find an nfa for the factor closure  ${_\sqsubseteq}L$,
  we make all non-empty states of the quotient automaton both accepting and initial
 and  delete the empty state. Hence the bound is $2^{n-1}$.
 The language $L$ defined by quotient automaton shown in Fig.~\ref{fig:suffix2}
 meets the bound.

 4. 
 To get an $\epsilon$-nfa for the subword-closure $ _{\Subset}L$ 
 from the quotient automaton of $L$,
 we remove the empty state
 (if there is no empty state, then  $_{\Subset}L=\Sigma^*$),
 and add an $\epsilon$-transition from state $p$ to state $q$
 whenever there is a transition from $p$ to $q$ in the quotient automaton.
 Since the initial state can reach every non-empty state
 through $\epsilon$-transitions,
 no other subset containing the initial state can be reached.
 Hence there are at most $2^{n-2}+1$  reachable subsets.

 To prove tightness, if $n=2$, let $\Sigma=\{a,b\}$; then $L=a^*$ meets the bound.
 If $n\ge 3$, let $\Sigma=\{a_1,\ldots,a_{n-2}\}$, and
 $L=\bigcup_{a_i\in\Sigma} a_i(\Sigma\setminus\{a_i\})^*.$
 Thus the language $L$ consists of all words over $\Sigma$,
 in which the first letter occurs exactly once.
 Let $K$ be the subword-closure of  $L$.
 Then $K=  L \cup \{w \in \Sigma^* \mid$
 at least one letter is missing in $w\}$.
 For each boolean vector $b=(b_1,b_2,\ldots,b_{n-2})$,
 define the word $w(b)=w_1w_2\cdots w_{n-2}$,
 in which $w_i=\epsilon$ if $b_i=0$ and $w_i=a_i$ if $b_i=1$.
 Now consider the   word $\epsilon$, and each word $a_1w(b)$.
 Let us show that all quotients of $K$
 by these $2^{n-2}+1$ words are distinct.
 For each binary vector $b$,
 we have $a_1a_2\cdots a_{n-2} \in K_\epsilon \setminus K_{a_1w(b)} $.
 Let $b$ and $b'$ be two different vectors with $b_i=0$ and $b'_i=1$.
 Then we have
 $a_1a_2\cdots a_{i-1}a_{i+1}a_{i+2}\cdots a_{n-2}
    \in K_{a_1w(b)} \setminus K_{a_1w(b')}$.
 Thus all quotients are distinct, and so $\kappa(K)\geq 2^{n-2}+1$.
\qed
\end{proof}

\section{Basic Operations on Closed Languages}
\label{***operations}

Now we study the quotient complexity of operations on closed languages. For regular languages, the following bounds are known~\cite{YZS94}: $mn$ for boolean operations, $m2^n-2^{n-1}$ for product, $3/4.2^n$ for star, and $2^n$ for reversal.
The bounds for closed languages are smaller in most cases. We also show that the bounds are tight, usually for a fixed alphabet.
The bounds for boolean operations and reversal follow from the results on ideal languages~\cite{BJL09}.

\begin{theorem}[Boolean Operations]\label{thm:BOOLEAN}
 If $K$ and $L$ are  prefix-closed (or factor-closed or subword-closed)
 with $\kappa(K)=m$ and $\kappa(L)=n$, then\\
 1. $\kappa(K\cap L) \le mn-(m+n-2)$,\\
 2.  $\kappa(K\cup L), \kappa(K\oplus L) \le mn$,\\
 3. $\kappa(K\setminus L)\le mn-(n-1)$,\\
 For suffix-closed languages, $\kappa(K\circ L)\leq mn$.
 All  bounds are tight if $|\Sigma|\geq 4$.
\end{theorem}

\begin{proof}
 Recall that the complement of a prefix-closed
 (respectively, suffix-, \mbox{factor-,} or subword-closed) language
 is a right (respectively, left, two-sided, all-sided) ideal.
 We get all the results using  De Morgan's laws and the results from~\cite{BJL09}.
\qed
\end{proof}
\begin{remark}\label{rem:prod}
 If $L$ is prefix-closed,
 then either $L=\Sigma^*$ or $L$ has $\emptyset$ as a quotient.
 Moreover, each quotient of $L$ is either accepting or $\emptyset$.
\end{remark}

\begin{remark}\label{rem:suffix}
 For a suffix-closed language $L$,
 if $v$ is a suffix of $w$ then $L_w\subseteq L_v$.
 In particular, $L_w\subseteq L_\epsilon=L$ for each word $w$ in $\Sigma^*$.
\end{remark}

\begin{theorem}[Product]\label{thm:PRODUCT}
 Let $K$ and $L$ be closed languages with $\kappa(K)=m$ and $\kappa(L)=n$,
 and let $k$ be the number of accepting quotients of $K$.
 If $m=1$ or $n=1$, then $\kappa(KL)=1$. Otherwise,\\
 1. If $K$ and $L$ are prefix-closed, then $\kappa(KL)\leq (m+1)\cdot2^{n-2}$.\\
 2. If $K$ and $L$ are suffix-closed, then $\kappa(KL)\leq (m-k)n+k$.\\
 3. If $K$ and $L$ are both factor- or both subword-closed,
    then   $\kappa(KL)\leq m+n-1$.\\
 All bounds are tight if $|\Sigma|\geq 3$.
\end{theorem}

\begin{proof}
 If $m=1$, then $K =\emptyset$ or $K=\Sigma^*$,
 and so $KL=\emptyset$ or, since $\epsilon\in L$, $KL=\Sigma^*$.
 Thus $\kappa(KL)=1$.
 The case of $n=1$ is similar.
 Now let $m,n\geq 2$. 

 1. If $K$ and $L$ are prefix-closed,
 then $\epsilon\in K$, and, by Remark~\ref{rem:prod},
 both languages have $\emptyset$ as a quotient.
 The quotient $(KL)_w$ is given by Equation~(\ref{eq:prod}).
 If $K_w$  is accepting, then $L$ is always in the union,
 and there are $2^{n-2}$ non-empty subsets of non-empty quotients of $L$
 that can be added.
 Since there are $m-1$ accepting quotients of $K$,
 there are $(m-1)2^{n-2}$ such quotients of $KL$.
 If $K_w$  is rejecting, then $2^{n-1}$ subsets of non-empty quotients of $L$
 can be added.
 Altogether, $\kappa(KL)\leq 2^{n-1}+ (m-1)2^{n-2}=(m+1)2^{n-2}$.

 For tightness,
 consider prefix-closed languages $K$ and $L$
 defined by the quotient automata of Fig.~\ref{fig:product_prefix}
 (if $n=2$, then $L=\{a,c\}^*$).
\begin{figure}[hbt]
    \centerline{\includegraphics[scale=0.40]{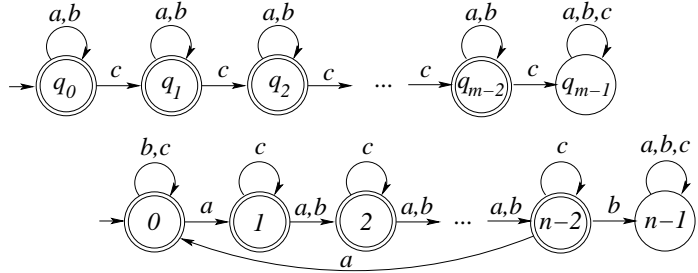}}
    \caption{Quotient automata
             of prefix-closed languages $K$ and $L$.}
    \label{fig:product_prefix}
 \end{figure}
 Construct an $\epsilon$-nfa
 for the language $KL$ from these quotient automata
 by adding an $\epsilon$-transition from states
 $q_0,q_1,\ldots,q_{m-2}$ to state $0$.
 The initial state of the nfa is $q_0$,
 and the accepting states are $0,1,\ldots,n-2$.
 Let us show that there are $(m+1)\cdot2^{n-2}$
 reachable and pairwise inequivalent states
 in the corresponding subset automaton.

 State $\{q_0,0\}$ is the initial state,
 and each state $\{q_0,0,i_1,i_2,\ldots,i_k\}$,
 where  $1\leq i_1 < i_2 < \cdots < i_k \leq n-2$,
 can be reached from state $\{q_0,0,i_2-i_1,\ldots,i_k-i_1\}$
 by word $ab^{i_1-1}$.
 For each subset $S$ of $\{0,1,\ldots,n-2\}$ containing state $0$,
 each state $\{q_i\}\cup S$ with $1\leq i \leq m-1$
 can be reached from state $\{q_0\}\cup S$  by $c^i$.
 If a non-empty set $S$ does not contain state $0$,
 then state $\{q_{m-1}\}\cup S$
 can be reached from state $\{q_{m-1}\}\cup \{s-\min S \mid s\in S\}$,
 which contains state $0$,  by $a^{\min S}$.
 State $\{q_{m-1},n-1\}$ can be reached from state $\{q_{m-1},n-2\}$ by $b$.

 To prove inequivalence, notice that the word $b^n$
 is accepted by the quotient automaton for $L$
 only from state $0$, and the word $a^{n-1-i}b^n$
 only from state $i$ ($1 \le i \leq n-2$).
 It turns out that two different states
 $\{q_{m-1}\}\cup S$ and $\{q_{m-1}\}\cup T$
 are inequivalent. It follows that states
 $\{q_{i}\}\cup S$ and $\{q_{i}\}\cup T$
 are inequivalent as well.
 States $\{q_{i}\}\cup S$ and $\{q_{j}\}\cup T$
 with $i<j$ can be distinguished by $c^{m-1-j}b^nab^n$.
 Hence the subset automaton has $(m+1)\cdot2^{n-2}$
 reachable and pairwise inequivalent states, and so
 $\kappa(KL)= (m+1)2^{n-2}$.

 2. If $K$ and $L$ are suffix-closed, then,
 by Remark~\ref{rem:suffix}, for each word $w$  we have
 \[
   (KL)_w=K_wL\cup K^\epsilon L_w
              \cup (\bigcup_{{w=uv}\atop {\;\; u,v\in\Sigma^+}}K^\epsilon_u L_v)
         =K_wL\cup L_x,
 \]
 for some suffix $x$ of $w$.
 If $K_w$ is a rejecting quotient,  there are at most $(m-k)n$ such quotients.
 If $K_w$ is accepting, then $\epsilon\in K_w$,
 and since $L_x\subseteq L_\epsilon= L\subseteq K_wL$,
 we have $(KL)_w=K_wL$.
 There are at most $k$ such quotients.
 Therefore there are at most $(m-k)n+k$ quotients in total.

 To prove tightness,
 let $K$ and $L$ be ternary  suffix-closed languages
 defined by quotient automata shown in Fig.~\ref{fig:product_suffix}.
 \begin{figure}[hbt]
    \centerline{\includegraphics[scale=0.40]{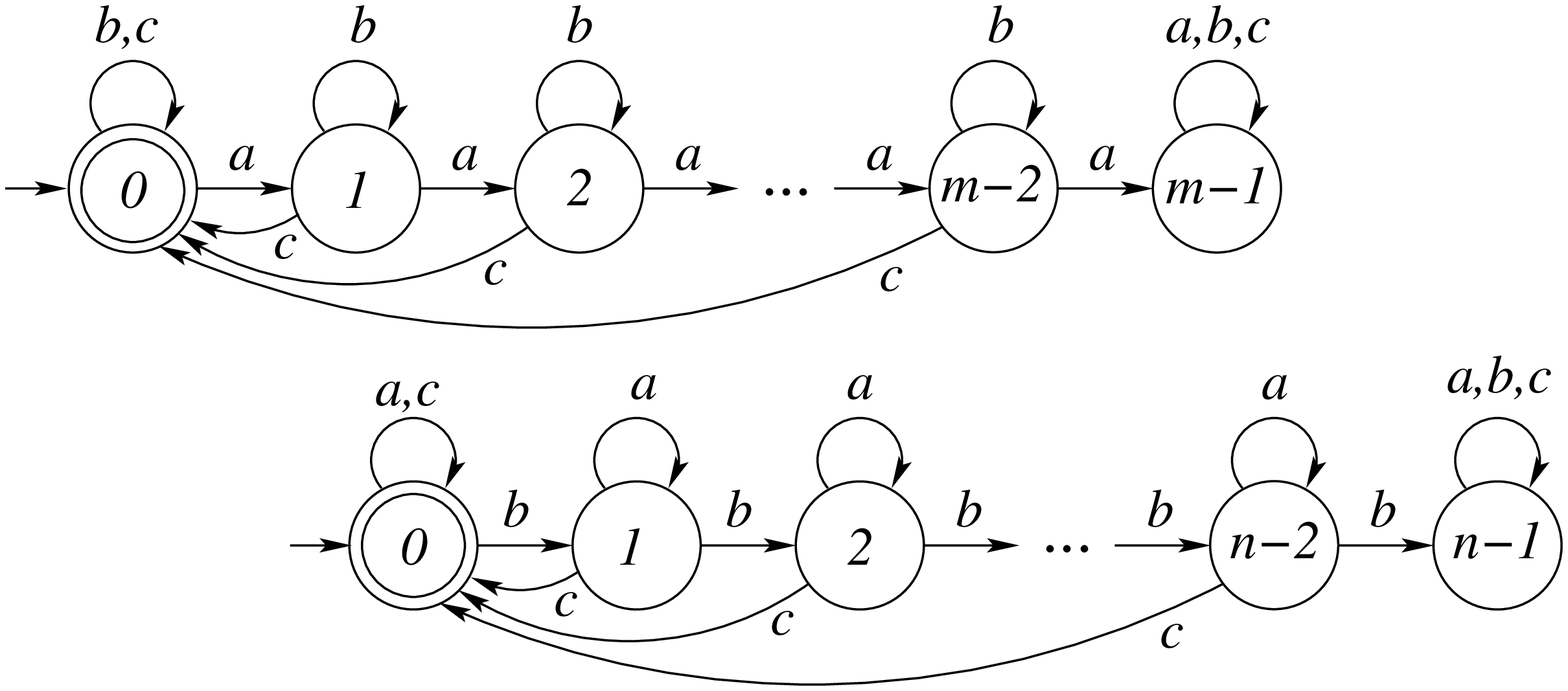}}
    \caption{Quotient automata
             of suffix-closed languages $K$ and $L$.}
    \label{fig:product_suffix}
 \end{figure}
 Consider the words $\epsilon=a^0b^0$, and $a^ib^j$
 with $1\leq i \leq m-1$ and $0\leq j \leq n-1$.
 Let us show that all quotients of  $KL$
 by these words are distinct.
 Let $(i,j)\neq(k,\ell)$, and let $x=a^ib^j$ and $y=a^kb^\ell$.
 If $i<k$, take $z=a^{m-1-k}b^{n}c$.
 Then $xz$ is in $KL$, while $yz$ is not, and so $z\in(KL)_x\setminus (KL)_y $.
 If $i=k$ and $j<\ell$, take $z=a^{m}b^{n-1-\ell}c$.
 We again have $z\in(KL)_x\setminus (KL)_y $.
 Thus the language $KL$ has at least $(m-1)n+1$ distinct quotients,
 and so $\kappa(KL)= (m-1)n+1$.

 Notice that, if the quotients $K_{a^i}$ with $0\leq i  \leq k-1$
 are accepting, then the resulting product
has quotient complexity $(m-k)n+k$.

 3.
 It suffices to derive the bound for factor-closed languages,
 since every subword-closed language is also factor-closed.
 Since factor-closed languages are suffix-closed,
 $\kappa(KL)\leq (m-k)n+k$.
 The language $K$ has at most one rejecting quotient,
 because it is prefix-closed.
 Thus, $k=m-1$ and $\kappa(KL)\leq m+n-1$.

 For tightness,
 consider binary subword-closed languages
 $K=\{w\in\{a,b\}^* \mid a^{m-1} \text{ is not a subword of } w\}$ and
 $L=\{w\in\{a,b\}^* \mid b^{n-1}$  is not a subword of $ w\}$
 with $\kappa(K)=m$ and $\kappa(L)=n$.
 Consider the word $w=a^{m-1}b^{n-1}$.
 This word is not in the product $KL$.
 However, removing any non-empty subword from $w$
 results in a word in  $KL$.
 Therefore, $\kappa(KL)\geq m+n-1$.
\qed
\end{proof}

\begin{theorem}[Star]\label{thm:STAR}
 Let $L$ be a closed language with $\kappa(L)=n\geq 2$.\\
 1. If $L$ is prefix-closed, then $\kappa(L^*)\leq 2^{n-2}+1$.\\
 2. If $L$ is suffix-closed, then $\kappa(L^*)\leq n$ if $L=L^*$
    and $\kappa(L^*)\leq n-1$ if $L\neq L^*$.\\
 3. If $L$ is factor- or subword-closed, then $\kappa(L^*)\leq 2$.\\
 If $\kappa(L)=1$, then $\kappa(L^*)\leq 2$.
 All bounds are tight if $|\Sigma|\ge2$.
\end{theorem}

\begin{proof}
1. For every non-empty word  $w$,
 the quotient $(L^*)_w$
 is given by Equation~(\ref{eq:star}).
 If $L$ is prefix-closed, then so is $L^*$ and $(L^*)_w$.
 Thus, if $(L^*)_w$ is non-empty,
 then it must contain the empty word.
 Hence $(L^*)_w\supseteq L^* \supseteq LL^*\supseteq L $.
 Since the empty quotient of $L$ and $L$ itself
 are always contained in every non-empty quotient of $L^*$,
 there are at most $2^{n-2}$ non-empty quotients of $L^*$.
 Since there is at most one empty quotient,
 there are at most $2^{n-2}+1$ quotients in total.
 The quotient $(L^*)_\epsilon$ has already been counted,
 since $L$ is closed and $\epsilon\in L$ implies $(L^*)_\epsilon=LL^*$,
 which has the form of Equation~(\ref{eq:star}).

 If $n=1$ and $n=2$,
 the bound 2 is met by $L=\emptyset$ and $L=\epsilon$,
 respectively.
 Now let $n\geq 3$
 and let $L$ be the prefix-closed language defined by  the quotient automaton
 shown in Fig.~\ref{fig:star_prefix}; 
 transitions not depicted in the figure go to state $n-1$.
\begin{figure}[hbt]
    \centerline{\includegraphics[scale=0.40]{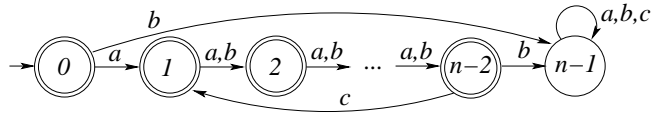}}
    \caption{Quotient automaton  of prefix-closed language $L$.}; 
    \label{fig:star_prefix}
 \end{figure}
 Construct an $\epsilon$-nfa  for   $L^*$
 by removing state $n-1$
 and adding an $\epsilon$-transition
 from all the remaining states to the initial state.
 Let us show that $2^{n-2}+1$ states are reachable
 and pairwise inequivalent
 in the corresponding subset automaton.

 We first prove that each subset of $\{0,1,\ldots,n-2\}$
 containing state 0 is reachable.
 The proof is by induction on the size of the subsets.
 The basis, $|S|=1$, holds true
 since $\{0\}$ is the initial state of the subset automaton.
 Assume that each set of size $k$ containing state 0
 is reachable, and let $S=\{0,i_1,i_2,\ldots,i_k\}$,
 where $0<i_1<i_2<\cdots<i_k\leq n-2$,
 be a set of size $k+1$.
 Then $S$ can be reached from the set
 $\{0,i_2-i_1,\ldots,i_k-i_1\}$ of size $k$ by $ab^{i_1-1}$.
 Since the latter set is reachable by the induction hypothesis,
 the set $S$ is reachable as well.
 The empty set can be reached from  $\{0\}$ by $b$,
 and we have $2^{n-2}+1$ reachable states.

 To prove inequivalence of these states
 notice that the word $b^{n-3}$
 is accepted by the nfa only from state $1$,
 and   each word $b^{n-2-i}cb^{n-3}$ ($2\leq i \leq n-2$),
 only from state $i$.
 It follows that all the states in the subset automaton
 are pairwise inequivalent.

 2. For a non-empty suffix-closed language $L$,
 the quotient $(L^*)_\epsilon$ is $LL^*$,
 which is of the same form as the quotients by a non-empty word $w$
 given by Equation~(\ref{eq:star}),
 $(L^*)_w=(L_w\cup L_{v_1}\cup \cdots \cup L_{v_k})L^*$,
 where the $v_i$ are suffixes of $w$, and $v_k$ is the shortest.
 By Remark~\ref{rem:suffix}, if $v$ is a suffix of $w$, then $L_w\subseteq L_v$.
 Thus the quotient becomes
 $(L^*)_w=L_{v_k}L^*$.
 There are at most $n$ such quotients.

 If $L\neq L^*$ for a non-empty suffix-closed language $L$,
 then there must be two words $x,y$ in $L$
 such that $xy\notin L$.
 Hence $y \in L_\epsilon \setminus L_x$,
 and so $L_\epsilon\neq L_x$.
 However, since $\epsilon\in L_x$ and
 $L^*$ is suffix-closed,
 we have $(L^*)_\epsilon=L^*\subseteq L_xL^*\subseteq (L^*)_x\subseteq (L^*)_\epsilon$,
 and so  $(L^*)_\epsilon=(L^*)_x$.
 It turns out that  $\kappa(L^*)\leq n-1$.

 For $n=1$, $L=\emptyset$ and for $n=2$, 
 $L=\epsilon$   meet the bound 2.
 Let $n\ge 3$.
 If $L=(a\cup ba^{n-2})^*$,
 then $L$ is suffix-closed, $\kappa(L)=n$, and $L^*=L$.
 If
 $L=\epsilon \cup \bigcup_{i=0}^{n-3} a^ib$,
 then $L$ is suffix-closed,  $\kappa(L)=n$,
 $L^*=(\bigcup_{i=0}^{n-3} a^ib)^*$,
 and $\kappa(L^*)=n-1$.

 3. If each letter in $\Sigma$ appears in some word
 of a factor-closed language $L$,
 then $L^*=\Sigma^*$ and $\kappa(L^*)=1$.
 Otherwise,  $\kappa(L^*)=2$.
 The bound is met by subword-closed language
 $L=\{w\in \{a,b\}^* \mid w=a^i \text{ and } 0\leq i\leq n-2\}$.
\qed
\end{proof}

Since the operation of reversal commutes with complementation,
we have the following results on ideal languages from~\cite{BJL09}:

\begin{theorem}[Reversal]\label{thm:REVERSAL}
 Let $L$ be a closed language with $\kappa(L)=n\geq 2$.\\
 1. If $L$ is prefix-closed, then $\kappa(L^R)\leq 2^{n-1}$.
     The bound is tight if $|\Sigma|\geq 2$.\\
 2. If $L$ is suffix-closed, then $\kappa(L^R)\leq 2^{n-1}+1$.
     The bound is tight if $|\Sigma|\geq 3$.\\
 3. If $L$ is factor-closed, then $\kappa(L^R)\leq 2^{n-2}+1$.
     The bound is tight if $|\Sigma|\geq 3$.\\
 4. If $L$ is subword-closed, then $\kappa(L^R)\leq 2^{n-2}+1$.
     The bound is tight if $|\Sigma|\geq 2n$.\\
 If   $\kappa(L)=1$, then $\kappa(L^R)=1$.
\qed
\end{theorem}

{\bf Unary Languages:} 
Unary closed languages have special properties
 because the product of unary languages is commutative.  
The classes of prefix-closed, suffix-closed, 
factor-closed, and subword-closed unary languages all coincide.
If a unary closed language $L$ is finite, 
then either it is empty and has $\kappa(L)=1$, 
or has the form $\{a^i\mid i\le n-2\}$, 
for some $n\ge 2$, and has $\kappa(L)=n$. 
If $L$ is infinite, then $L=a^*$, and $\kappa(L)=1$.
The bounds for unary languages 
are  given in Tables~\ref{tab:bool} and~\ref{tab:gen}
on page~\pageref{tab:bool}.

\section{Kuratowski Algebras Generated by Closed Regular  Languages}
\label{***kuratowski}

A theorem of Kuratowski~\cite{Kur22} states that,
given a topological space,
at most 14 distinct sets can be produced
by repeatedly applying the operations
of closure and complement to a given set.
A closure operation on a set $S$
is an operation $\Box:2^{S}\rightarrow2^{S}$
satisfying the following conditions for any subsets $X,Y$ of $S$:
(1)~$X\subseteq X^\Box$,
(2) $X\subseteq Y$ implies $X^\Box\subseteq Y^\Box$,
(3) $X^{\Box\Box}\subseteq X^\Box$.

Kuratowski's theorem was studied
in the setting of formal languages in~\cite{BGS09}.
Positive closure and Kleene closure (star) are both closure operations.
It was shown in~\cite{BGS09} that at most 10 distinct languages
can be produced by repeatedly applying the operations
of positive closure and complement to a given language,
and at most 14 distinct languages can be produced
with Kleene closure instead of positive closure.
We consider here the case where the given language is closed and regular,
and give upper bounds for the complexity of the resulting languages.
Here we denote the complement of a language $L$ by $L^-$.
Moreover, the positive closure of the complement of $L$ is denoted by $L^{-+}$, \etc\/

We begin with positive closure.
Let $L$ be a $\unlhd$-closed language not equal to $\Sigma^*$.
Then $L^-$ is an ideal, and $L^{-+}=L^-$.
In addition, $L^+$ is also $\unlhd$-closed, so $L^{+-+}=L^{+-}$.
Hence there are at most 4 distinct languages
that can be produced with positive closure and complementation.

\begin{theorem}\label{thm:POSITIVE}
 The worst-case complexities in every 4-element algebra
 generated by a closed language $L$ with $\kappa(L)=n$ 
 under positive closure and complement are:
 $\kappa(L)=\kappa(L^-)=n$,
 $\kappa(L^+)=\kappa(L^{+-})=f(n)$,
 where $f(n)$ is:
 $2^{n-2}+1$ for prefix-closed languages,
 $n-1$ for suffix-closed languages, and
 $2$ for factor- and subword-closed languages.
 There exist closed languages that  meet these bounds.
\end{theorem}

\begin{proof}
 Since $L^+=L^*$ for a non-empty closed language
 we have $\kappa(L^+)= \kappa(L^*)$,
 and the upper bounds $f(n)$ follow
 from our results on the quotient complexity of  star operation;
 in the case of suffix-closed languages,
 to get a 4-element algebra we need  $L\neq L^*$.
 All the languages that we have used in Theorem~\ref{thm:STAR}
 to prove tighness can be used as examples meeting the bound $f(n)$.
\qed
\end{proof}

The case of Kleene closure is similar.
Let  be a $\unlhd$-closed language such that
$L\not\in\{\emptyset,\Sigma^*\}$.
Then $L^-$ is an ideal and $L^-$ does not contain $\epsilon$.
Thus $L^{-*}=L^-\cup\epsilon$ and $L^{-*-}=L\setminus \epsilon$,
which gives at most four languages thus far.
Now $L^*=(L\setminus\epsilon)^*$,
and $L^*$ is also $\unlhd$-closed.
By the previous reasoning, we have at most four additional languages,
giving a total of eight languages as the upper bound.
The 8-element algebras are of the form
$
(L,\,L^-,\,L^{-*}=L^-\cup\epsilon,\,L^{-*-}=L\setminus\epsilon,\,L^*,\,L^{*-},\,L^{*-*}=L^{*-}\cup\epsilon,\,L^{*-*-}=L^*\setminus\epsilon)
.$
\begin{theorem}\label{thm:KLEENE}
 The worst-case complexities in
 every 8-element algebra generated 
 by a closed language $L$ with $\kappa(L)=n$ 
 under Kleene closure and complement are:
 $\kappa(L)=\kappa(L^-)=n$,
 $\kappa(L^*)=\kappa(L^{*-})=f(n)$,
 $\kappa(L^{*-*})=\kappa(L^{*-*-})=f(n)+1$,
 $\kappa(L^{-*})=\kappa(L^{-*-})=n+1$,
 where $f(n)$ is:
 $2^{n-2}+1$ for prefix-closed languages,
 $n-1$ for suffix-closed languages, and
 $2$ for factor-and subword-closed languages,
 Moreover, there exist closed languages that  meet these bounds.
\end{theorem}

\begin{proof}
 Since $L^{-*-}=L\setminus\epsilon$ and $L^{*-*-}=L^*\setminus\epsilon$
 we have
 $\kappa(L^{-*-})\le n+1$ and $\kappa(L^{*-*-})\le f(n)+1$.
 In the case of suffix-closed languages, since $L$ must be distinct from $L^*$,
 we have $f(n)=n-1$ by Theorem~\ref{thm:STAR}.

 1. Let $L$  be the prefix-closed language defined
 by the quotient automaton in Fig.~\ref{fig:star_prefix}
 on page~\pageref{fig:star_prefix}; then $L$
meets the upper bound on star.  Add a loop with a new letter $d$ in each state
 and denote the resulting language by $K$.
 Then $K$ is a prefix-closed language with
 $\kappa(K)=n$ and $\kappa(K\setminus\epsilon)=n+1$.
 Next we have
 $\kappa(K^*)=\kappa(L^*)=2^{n-2}+1$ and $\kappa(K^*\setminus\epsilon)=2^{n-2}+2$.

 2. Let $L=b^*\cup \bigcup_{i=1}^{n-3}b^*a^ib$.
 Then $L$   is a suffix-closed language with
 $\kappa(L)=n$ and $\kappa(L\setminus\epsilon)=n+1$.
 Next,  $\kappa(L^*)=n-1$, and $\kappa(L^*\setminus\epsilon)=n$.

 3. Let $L=\{w\in \{a,b,c\}^* \mid w=b^*a^i \text{ and } 0\leq i \leq n-2\}$.
 Then $L$ is a subword-closed language
 with $\kappa(L)=n$ and $\kappa(L\setminus\epsilon)=n+1$.
 Next $L^*=\{a,b\}^*$, and so $\kappa(L^*)=2$ and $\kappa(L^*\setminus\epsilon)=3$.
\qed
\end{proof}

\section{Conclusions}
\label{***conclusions}
Tables~\ref{tab:bool} and~\ref{tab:gen} summarize our complexity results.
The complexities for regular languages are from~\cite{YZS94}, except those for difference and symmetric difference, which are from~\cite{Brz09}. The bounds for boolean operations and reversal of closed languages are direct consequences of the results in~\cite{BJL09}.
In Table~\ref{tab:gen}, $k$ is the number of accepting quotients of $K$.
\begin{table}[ht]
\begin{center}
$
\begin{array}{|| l || c | c | c | c || }
\hline
\hline
 &  K\cup L& K\cap L    &   K\setminus L
&  K\oplus L
\\
\hline
\hline
\txt{unary closed}
&   max(m,n) & \,max(m,n) \,
& m
& max(m,n)
\\
\hline
\hline
\txt{$\leq$-, $\sqsubseteq$-, $\Subset$-closed}
&    mn & mn-(m+n-2)   &   mn-(n-1)
&  mn     \\
\hline
\txt{$\preceq$-closed}
&     mn & mn   &  mn
&  mn   \\
\hline
\hline
\txt{regular     }
&mn&mn &mn    &mn\\
\hline
\hline
\end{array}
$
\end{center}
\caption{Bounds on quotient complexity of boolean operations.}
\label{tab:bool}
\end{table}
\vskip-20pt
\begin{table}[ht]
\begin{center}
$
\begin{array}{|| l ||c|c|c|c||} \hline
\hline
 &  _\unlhd L  & KL &   K^*
 & \, K^\rhoR \,
 \\
\hline
\hline
\txt{unary closed}
&  n  &   m+n-2    &2 & n\\
\hline
\hline
\txt{$\leq$-closed}
&  n  &    m2^{n-2} & 2^{n-2}+1 & 2^{n-1}\\
\hline
\txt{$\sqsubseteq$-closed}
&\;  2^{n-1}\;  &  m+n-1  & 2 & 2^{n-2}+1 \\
\hline
\txt{$\Subset$-closed}
&  2^{n-2}+1 & m+n-1 & 2 & 2^{n-2}+1\\
\hline
\txt{$\preceq$-closed}
&  2^n-1    &  (m-k)n+k & n & 2^{n-1}+1\\
\hline
\hline
\txt{regular}
& -  &\;m2^n-k2^{n-1}\;  & \;2^{n-1} + 2^{n-k-1} \; & 2^n\\
\hline
\hline
\end{array}
$
\end{center}
\caption{Bounds on quotient complexity of closure, product, star and reversal.}
\label{tab:gen}
\end{table}
\medskip

\bibliographystyle{splncs}

\end{document}